\documentclass[aps,pra,twocolumn]{revtex4-1}
\usepackage{colordvi,epsfig,color,amsmath,amssymb}

\def\be{\begin{equation}} 
\def\ee{\end{equation}} 
\def\bee{\begin{eqnarray}} 
\def\eee{\end{eqnarray}}

\def\kb{k_{\rm B}}

\def\halb{\mbox{$\frac{1}{2}$}}

\newcommand{\bbbone}{{\mathchoice {\rm 1\mskip -4mu l}{\rm 1\mskip -4mu l}{\rm
1\mskip -4.5mu l}{\rm 1\mskip -5mu l}}}

\begin{document}

\title{Crossing time in the Landau-Zener quantum dynamics in a super Ohmic environment}

\author{P. Nalbach$^{1}$}
\affiliation{
$^1$Fachbereich Wirtschaft \& Informationstechnik, Westf\"alische Hochschule, M\"unsterstrasse 265, 46397 Bocholt, Germany\\
}

\date{\today}

\begin{abstract}

We study the dynamics of a quantum two state system driven through an avoided crossing under the influence of a super Ohmic environment, i.e. a longitudinal as well as a transversal one. The crossing time window, in which relaxation influences the dynamics, is centered around the avoided crossing. 
We determine the dynamics and the Landau-Zener probability employing the numerical exact quasi-adiabatic path integral. At weak coupling we show that the numerically less demanding nonequilibrium Bloch equations provide an accurate description. The crossing time depends strongly not only on the system-bath coupling strength but also on the bath spectral cut-off frequency in contrast to the situation in an Ohmic bath. Our results enable to design quantitative protocols which drive quantum systems out of the influence range of relaxation.

\end{abstract}


\maketitle

\section{Introduction}

Dissipation of a quantum system is determined by its interaction with an environment and its spectrum \cite{WeissBuch}. Accordingly, for a driven quantum system dissipation changes with time due to the external control fields. Coherent control of quantum systems is at the heart of quantum devices and dissipation the main obstacle to realize them \cite{Castelvecchi2017, Herbschleb2019, Miao2020}. Thus, a clear understanding of the interplay of driving and dissipation is key for future quantum electronics \cite{KeckNJP2017, CarregaNJP2020, BurkardPRB2020}.

A quantum two-state system (TSS) (linearly) driven through an avoided crossing (Landau-Zener dynamics) \cite{Landau1932, Zener1932, Majorana1932, Stueckelberg1932} is a widely employed theoretical model to study dissipation in driven quantum systems \cite{KayanumaLZ1984a, KayanumaLZ1984b, AoLZ1989, KayanumaLZ1998, WubsLZ2006, PokrovskyLZ2007, NalbachLZ2009, NalbachLZ2010, NalbachLZ2015, WhitneyPRL2011, ZhaoPRA2018, RaikhPRB2018, ChenPRB2020}.
Typically, the Landau-Zener probability $P_{\rm LZ}$ to end up in the ground state when the system started there or, alternatively, the excited state survival probability $P_{\rm ES}$ to end up in the excited state when the system started from the excited state is studied. While for nonadiabatic driving an environment hardly influences the Landau-Zener dynamics, for adiabatic driving a crossover is observed where at finite temperature relaxation dominates the Landau-Zener dynamics \cite{NalbachLZ2009, NalbachLZ2010, NalbachLZ2015}. 

The Landau-Zener dynamics is determined by the ratio between the driving and the tunnel coupling between the two states. The protocol runs for an infinite time but the tunnel coupling influences the dynamics only while it exceeds the energy splitting. This allows to define the (coherent) Landau-Zener crossing time during which the avoided crossing takes place. In contrast, one expects relaxation to drive the TSS always towards thermal equilibrium and, accordingly, the Landau-Zener probability to approach unity when the TSS is coupled to an environment. Since driving modifies also the system-bath interaction, this is not the case. A longitudinal Ohmic bath influences the Landau-Zener probability only within a time window similar to the (coherent) Landau-Zener crossing time \cite{NalbachLZ2009, NalbachLZ2010}. A transversal Ohmic bath, in contrast, allows relaxation in a much wider time window around the avoided crossing \cite{NalbachLZ2015} determined by temperature and the environmental spectrum. 

Since the crossing time window, in which dissipation is relevant, depends strongly on the environment, we extend the studies so far by focusing on a super Ohmic bath. The influence of a super Ohmic environment on the Landau-Zener dynamics has not been studied in detail yet. Whitney et al. \cite{WhitneyPRL2011} studied super Ohmic influence on the coherent oscilations in the dynamics after the avoided crossing.  
We observe that a longitudinal super Ohmic bath shows a crossing time window similar to a transversal Ohmic bath. A transverse super Ohmic bath, however, exhibits again a much wider crossing time window which renders a numerical exact simulation even at very small system-bath couplings challenging. We determine the Landau-Zener and the excited state survival probabilities numerically exact employing the quasi adiabatic path integral (QUAPI) \cite{MakriQUAPI1995a, MakriQUAPI1995b, NalbachLZ2009, PalmJCP2018}. We show that at weak coupling the adiabatic Markovian non-equilibrium Bloch equations (NEQBs) \cite{NalbachLZ2013, NalbachLZ2014} can reliably simulate both. This emphasizes that relaxation dominates and dephasing is less effective in influencing the Landau-Zener dynamics since a Markovian approximation in a super Ohmic bath is not expected to treat dephasing in an asymmetric TSS correctly \cite{WeissBuch}. The crossing time windows can directly be rationalized by the drive dependent weak coupling relaxation rates of the NEQBs. In the next three sections we introduce our model, our observables and the two methods, i.e. QUAPI and NEQB. In section \ref{chapLZdynamics} we determine the Landau-Zener and the excited state survival probability as function of drive parameters for various temperatures and system-bath couplings. Having validated the usage of the NEQB at weak system bath coupling we discuss the crossing time windows in which relaxation actively influences the dynamics. Finally, we end with a conclusion.

\section{Model}\label{SecModel}

A quantum two-level system (TSS) driven through an avoided crossings is modelled by an Hamiltonian
\be\label{eq1} H_S(t) \,=\, \frac{\Delta}{2}\sigma_x \,+\,\frac{\epsilon(t)}{2}\sigma_z \, 
\ee
with Pauli matrices $\sigma_i$, driving field $\epsilon(t)$ and tunnel coupling $\Delta$ between the two eigenstates of the system at $t=\pm\infty$. We discuss a linear driving protocol, i.e. $\epsilon(t)=vt$ with sweep or drive speed $v$ and the avoided crossing at $t=0$.

We model environmental noise acting on the TSS in a system-bath approach \cite{WeissBuch, Leggett1987} by coupling a bath of harmonic oscillators, i.e. $H_B=\sum_k\omega_k b_{k}^\dagger b_{k}$, bilinearly to the TSS resulting in a total Hamiltonian \cite{WeissBuch, Leggett1987}
\be\label{hges} H(t) \,=\, H_S(t) - \frac{\sigma_z\cos\Theta + \sigma_x\sin\Theta }{2}\hat{B}  + H_{B} 
\ee
and $\hat{B}=\sum_k\lambda_{k} (b_{k}+b_{k}^\dagger)$ 
with bosonic annihilation $b_{k}$ and creation $b_{k}^\dagger$ operators. The system-bath coupling is denoted as {\it longitudinal} for $\Theta=0$ and {\it transversal} for $\Theta=\pi/2$. The bath influence is captured by a spectral function
\be
G(\omega)=\sum_k \lambda_{k}^2 \delta(\omega-\omega_k)=\frac{\gamma\Delta^{1-s}}{\pi}\omega^s \exp(-\omega/\omega_{c})
\ee
with cut-off frequency $\omega_{c}$, the coupling strength $\gamma$ and the spectral exponent $s$. For $s=1$ the bath is termed \textit{Ohmic} and for $s>1$ \textit{super-Ohmic}. In the following we focus on $s=3$ which reflects, for example, phonons in a 3D system coupled to an acoustic dipole \cite{Wuerger1997, Wuerger1998, NalbachRESPET2002, NalbachLudwigTS2003, NalbachJLTPTS2004}.

\section{Observables}

In the Landau-Zener protocol we drive the TSS with $\epsilon(t)=vt$ starting at $t=-\infty$ and ending at $t=\infty$. We determine the Landau-Zener probability $P_{\rm LZ}$ to end up in the ground state when the system started in the ground state, i.e.
\be 
P_{\rm LZ} = \mbox{Tr}\{ \hat{g} \rho(t=\infty) \}
\ee
with $\hat{g}=\halb (\bbbone-\sigma_z)$ and $\rho(t=\infty)$ the statistical operator at the end of the driving protocol when $\rho(t=-\infty)=\hat{g}$. We also calculate the excited state survival probability $P_{\rm ES}$ to end up in the excited state when the system started from the excited state, i.e. 
\be 
P_{\rm ES} = \mbox{Tr}\{ \hat{e} \rho(t=\infty) \}
\ee
with $\hat{e}=\halb (\bbbone+\sigma_z)$ and $\rho(t=\infty)$ the statistical operator at the end of the driving protocol when $\rho(t=-\infty)=\hat{e}$.

Without environment (indicated by the superscript $(0)$) these probabilities are
\be 
P_{\rm LZ}^{(0)} = P_{\rm ES}^{(0)} = 1 - e^{-\frac{\pi\Delta^2}{2v}}
\ee
with adiabatic behaviour for $v \ll \Delta^2$, diabatic behaviour for $v \gg \Delta^2$ and a mixed regime for $v\simeq \Delta^2$ \cite{Landau1932, Zener1932, Majorana1932, Stueckelberg1932}. 

\section{Methods}

Including now the coupling to the bath, closed forms for $P_{\rm LZ}$ and $P_{\rm ES}$ are not available and we are forced to determine the time dependent dynamics, i.e. the reduced density $\rho(t)$ of the TSS, numerically. The Landau-Zener driving protocoll runs from the infinite past to the avoided crossing at $t=0$ to the infinite future. Numerical investigation relies on the fact that the dynamics is determined in the proximity of the avoided crossing. Accordingly, converged results are quickly obtained for protocols running from $-t_{\rm max}/2$ to $t_{\rm max}/2$ when increasing $t_{\rm max}$. An extended time window in which relaxation is finite, renders the Landau-Zener dynamics supstantially more difficult to study numerically since with increasing investigation time $t_{\rm max}$ the numerically accumulated errors increase.

\subsection{Nonequilibrium Bloch equations}

For weak system-bath coupling a Markovian Redfield type approach is known to describe the Landau-Zener dynamics correctly \cite{HaenggiPRE2000, PekolaPRB2010, NalbachLZ2013, NalbachLZ2014, SantoroLZ2017, SantoroLZ2017Erratum} for an Ohmic bath. It leads to relaxation and dephasing rates with time  dependence governed by the time dependence of the driving field. We employ the nonequilibrium Bloch equations (NEBQs) \cite{NalbachLZ2014}. Diagonalizing the system (Eq.(\ref{eq1})) to $H_S(t)=E(t)\tau_x/2$ with Pauli matrices $\tau_i$ and $E(t)=\sqrt{\Delta^2+\epsilon^2(t)}$, and introducing the reduced density operator for the system $\rho_S(t)=\halb(\bbbone-r_i\tau_i)$ the NEQBs are
\be\label{blocheq} 
\begin{array}{lcrrl}
\partial_t r_x(t) & = &     +\phi'(t) r_z(t) &                       & \hspace*{-1.5cm}-\Gamma_1(t) [r_x(t)-r_x^{st}(t)]  \\
\partial_t r_y(t) & = &  -\Gamma_2(t) r_y(t) &         -E(t) r_z(t)                                                     \\
\partial_t r_z(t) & = &         +E(t) r_y(t) &  -\Gamma_2(t) r_z(t)  &                  -\phi'(t) r_x(t)  
    \end{array}
\ee
with $\phi(t)=\arctan(\epsilon(t)/\Delta)$, $r_x^{(st)}(t)=\tanh(\beta E(t)/2)$ and 
\bee
\Gamma_{1}[s,\Theta](t) &=&  A_\Theta^2(t) \frac{\pi}{2} G(E(t)) \coth\left( \frac{\beta E(t)}{2}\right) 
\eee
with
\bee
A_\Theta(t) &=& u(t)\cos\Theta - v(t)\sin\Theta \nonumber\\
u(t) &=& \cos\varphi(t)=\frac{\Delta}{E(t)} \nonumber\\
v(t) &=& \sin\varphi(t)=\frac{\epsilon(t)}{E(t)} . \nonumber
\eee
and 
\be
\Gamma_2[s,\Theta](t) ´= \frac{1}{2} \Gamma_{1}[s,\Theta](t) + \Gamma_d[s,\Theta](t) 
\ee
with
\bee
\Gamma_d[s,\Theta] &=& B_\Theta^2(t) \frac{\pi}{2} \lim_{\omega\to 0} \bigl[ G(\omega) \coth(\beta\omega/2) \bigr] \\
&=& \left\{ 
\begin{array}{lcl}
B_\Theta^2(t) \gamma \kb T & : & s=1 \\ 0 & : & s > 1
\end{array} \right. \quad \mbox{with} \nonumber \\
 B_\Theta(t) &=& v(t)\cos\Theta + u(t)\sin\Theta . \nonumber
\eee
The pure dephasing rate $\Gamma_d(t)$ has only for an Ohmic environment a finite Markovian contribution. However, for an asymmetric TSS (which exhibits a finite $B_\Theta$) in a super Ohmic bath the Markovian approximation is not reliable for describing pure dephasing \cite{WeissBuch}. 

\subsection{QUAPI}

The quasi-adiabatic path integral (QUAPI) \cite{MakriQUAPI1995a, MakriQUAPI1995b, NalbachLZ2009, PalmJCP2018} was used to obtain numerical exact results for the dissipative Landau Zener dynamics. 
QUAPI is based on a symmetric Trotter splitting of the short-time propagator ${\cal K}(t_{k + 1}, t_k)$ (describing time evolution over a time slice $\delta t$) for the full Hamiltonian $H$. The splitting is by construction exact in the limit $\delta t \to 0$, but introduces a finite Trotter error for a finite time increment, which has to be eliminated by choosing $\delta t$ small. The QUAPI scheme further employs an approximated Feynman-Vernon influence functional which includes only non local time correlations between observables in a time window $\tau_{\rm mem}=k_{\rm max} \delta t$. Thus, valid results are achieved by finding convergence while increasing $\tau_{\rm mem}$ but at the same time decreasing $\delta t$ to minimize the Trotter error. In the following, only converged results are presented and discussed. For a discussion on how to achieve convergence see \cite{MakriQUAPI1995b, PalmJCP2018}.

\section{Landau-Zener Dynamics}\label{chapLZdynamics} 

\subsection{Longitudinal super Ohmic bath $\Theta=0$}

\begin{figure}[t]
\includegraphics[width=8.5cm]{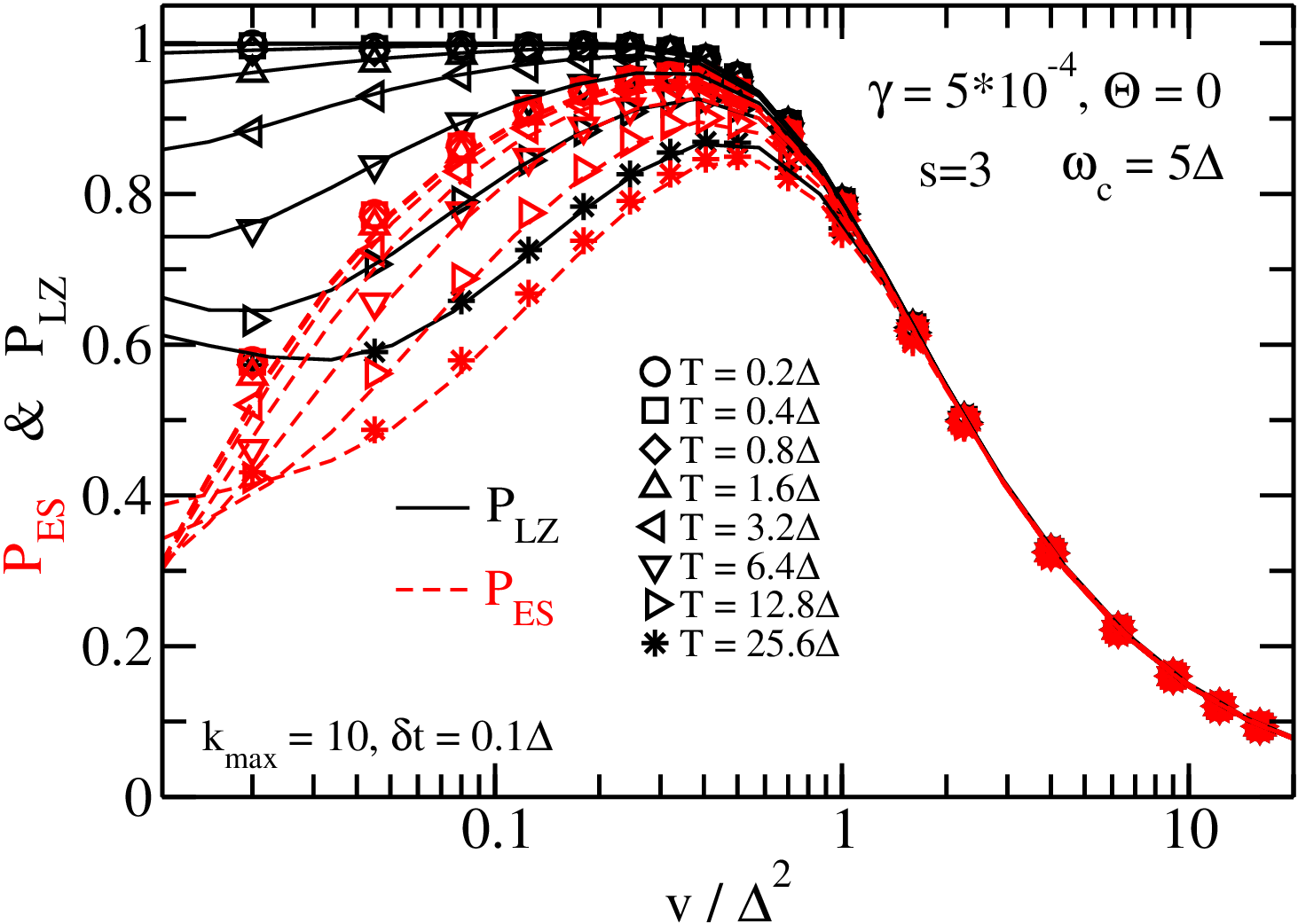}
\caption{\label{Fig1Nal} The Landau-Zener probability $P_{\rm LZ}$ and the excited-state-survival probability $P_{\rm ES}$ versus sweep speed for various temperatures at $\omega_c=5\Delta$, $\gamma=5\cdot 10^{-4}$ and $\Theta=0$.}
\end{figure}

Fig. \ref{Fig1Nal} presents results for $P_{\rm LZ}$ (full black lines) and $P_{\rm ES}$ (dashed red lines) versus sweep speed for various temperatures for a system-bath coupling strength $\gamma=5\cdot 10^{-4}$ to a longitudinal super-Ohmic bath. QUAPI results are given as symbols and lines are NEQB results. Numerical uncertainties are roughly given by the symbol size. QUAPI and NEQB results agree well, thus, justifying the weak coupling NEQB approach. The Landau-Zener probability is not influenced by dissipation at low temperatures $\kb T\lesssim \Delta$ nor for fast drives $v\gtrsim \Delta^2$. For $\kb T >\Delta$ and $v < \Delta^2$ the Landau-Zener probability evolves a minimum which deepens with increasing temperature. The drive speed $v_{\rm min}$ at the minimum is roughly determined by the condition that the time within the crossing time window equals the (averaged) relaxation time itself \cite{NalbachLZ2010}. The picture here is that too slow relaxation does not influence the dynamics but too fast relaxation would only bring the system back to the ground state of the TSS whose energy splitting increases continuously after the crossing.
Similar qualitative and quantitative results have been obtained for a transversal Ohmic bath \cite{NalbachLZ2015} at the same system-bath coupling strength and for a longitudinal Ohmic bath with $\gamma$ an order of magnitude larger. 

\begin{figure}[t]
\includegraphics[width=8.5cm]{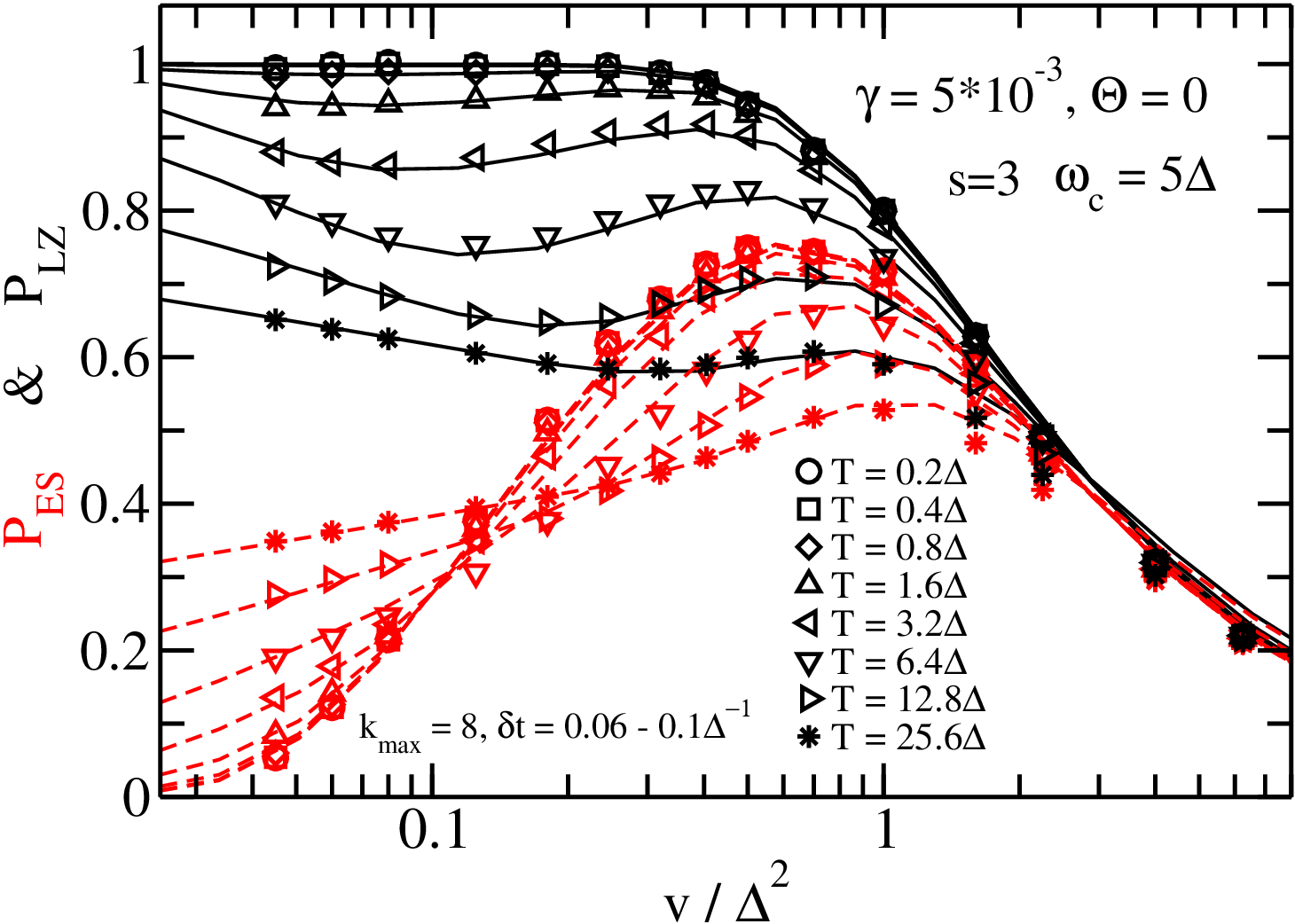}
\caption{\label{Fig2Nal} The Landau-Zener probability $P_{\rm LZ}$ and the excited-state-survival probability $P_{\rm ES}$ versus sweep speed for various temperatures at $\omega_c=5\Delta$, $\gamma=5\cdot 10^{-3}$ and $\Theta=0$.}
\end{figure}

For a stronger system-bath coupling $\gamma=5\cdot 10^{-3}$ (see Fig. \ref{Fig2Nal}) QUAPI and NEQB results agree well in the adiabatic regime. For $v\gtrsim \Delta^2$ at higher temperatures $T\gtrsim 3\Delta/\kb$ NEQB starts overestimating the probabilities $P_{\rm LZ}$ and $P_{\rm ES}$. With increasing system bath coupling the sweep speed of the minimum $v_{\rm min}$ is shifted to larger $v$. 

For even larger system bath coupling QUAPI convergence is hard to achieve within reasonable numerical effort and QUAPI and NEQB differ in the whole range of $v$ values studied. Besides QUAPI convergence with its parameters $k_{\rm max}$ and $\delta t$ we must ensure convergence of the driving protocol, i.e. in regard to the time window parameter $t_{\rm max}$ for the simulation running from $-t_{\rm max}/2$ to $t_{\rm max}/2$. Roughly, in the presented data $t_{\rm max}$ varies between $50\Delta^{-1}$ for $v\gg\Delta^2$ to $1500\Delta^{-1}$ for $v\ll\Delta^2$ depending also weakly on temperature. Generally, $P_{\rm ES}$ needs larger $t_{\rm max}$ compared to $P_{\rm LZ}$.

\subsection{Transversal Bath $\Theta=\pi/2$}

\begin{figure}[t]
\includegraphics[width=8.5cm]{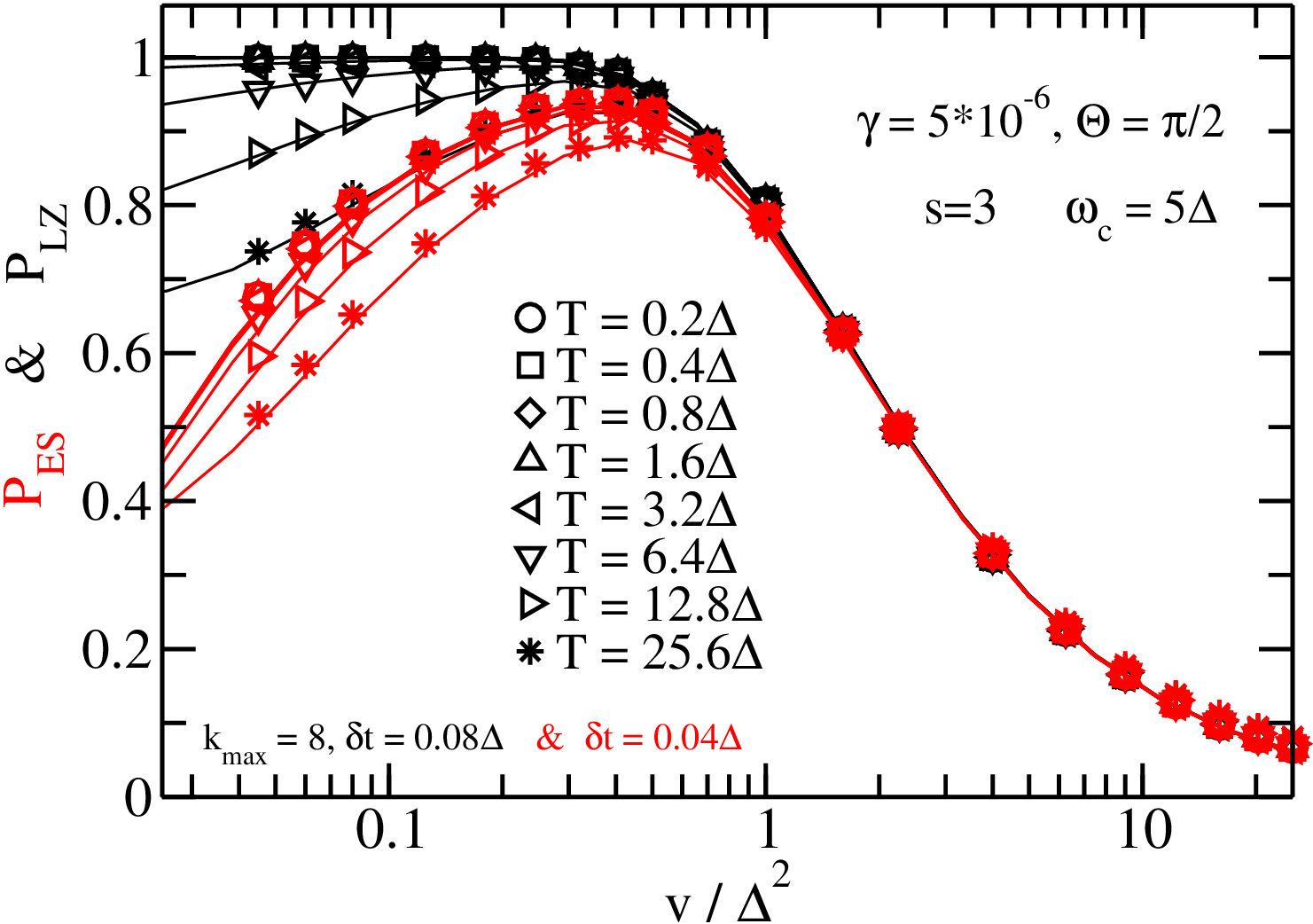}
\caption{\label{Fig3Nal} The Landau-Zener probability $P_{\rm LZ}$ and the excited-state-survival probability $P_{\rm ES}$ versus sweep speed for various temperatures at $\omega_c=5\Delta$, $\gamma=5\cdot 10^{-6}$ and $\Theta=\pi/2$.}
\end{figure}

In case of a transversal super Ohmic bath influencing the Landau Zener system we are only able to achieve convergence for a very small system bath coupling strength of $\gamma=5\cdot 10^{-6}$.
Fig. \ref{Fig3Nal} presents results for $P_{\rm LZ}$ and $P_{\rm ES}$ versus sweep speed for various temperatures. QUAPI results are given as symbols and lines are NEQB results. Both agree well, thus, justifying a weak coupling NEQB approach. Roughly, similar dynamics is observed for a longitudinal bath for coupling strength two orders of magnitude larger.

Converged results for very small sweep speeds $v$ or larger system bath couplings could not be achieved within reasonable numerical effort. In Fig. \ref{Fig3Nal} the minimum in $P_{\rm LZ}$ is not resolved but will be found at smaller sweep speeds according to the easily obtained NEQB results. Thus, convergence is not hindered by long time memory due to strong system bath coupling but merely due to the very long simulation times needed for convergence regarding $t_{\rm max}$. 

\subsection{Mixed Bath $-\pi/2<\Theta\le\pi/2$}

To complete our investigation of the Landau Zener dynamics in a super Ohmic environment we have additionally studied mixed bath cases, i.e. a single bath which exhibits both, longitudinal and transversal coupling with $-\pi/2<\Theta\le\pi/2$. For an Ohmic bath it was observed that small transversal contributions to a mainly longitudinal bath shift the minimum in $P_{\rm LZ}$, i.e. $v_{\rm LZ}$, depending on the sign of the angle $\Theta$. Arising from this asymmetry an environmental rocking ratchet can be designed which allows rectification in energy transport through an according system \cite{NalbachPRE2017}.

\begin{figure}[t]
\includegraphics[width=8.5cm]{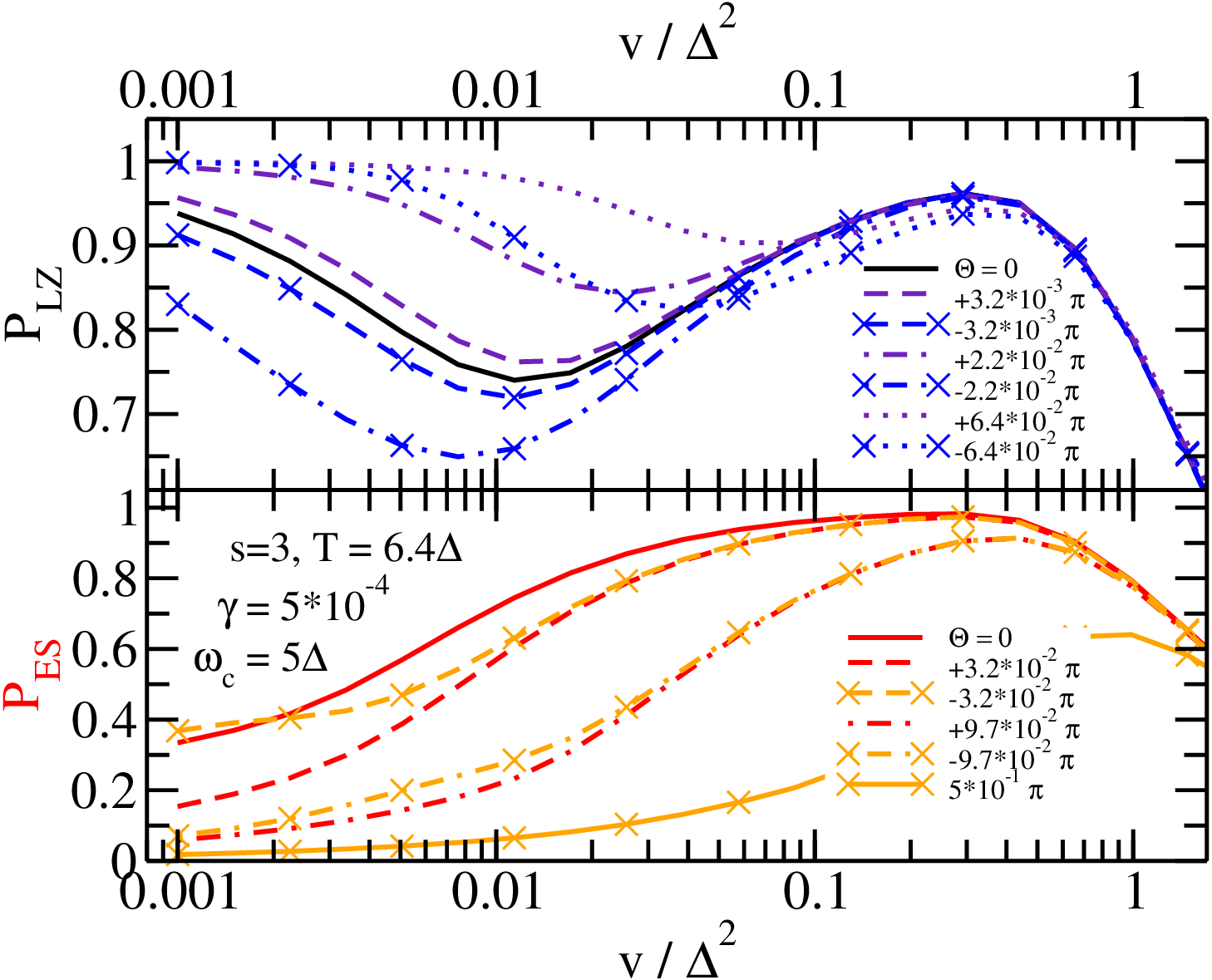}\caption{\label{Fig4Nal} The Landau-Zener probability $P_{\rm LZ}$ (upper figure) and the excited-state survival probability $P_{\rm ES}$ (lower figure) versus sweep speed for various mixing angles $-\pi/2<\Theta\le\pi/2$ at fixed temperature $T=6.4\Delta/\kb$ and $\omega_c=5\Delta$ and $\gamma=5\cdot 10^{-4}$.}
\end{figure}

Fig. \ref{Fig4Nal} presents the Landau Zener probability (upper figure) and the excited state survival probability (lower figure) versus drive speed $v$ for various mixing angles $\Theta$, at fixed temperature $T=6.4\Delta/\kb$ and a system bath coupling $\gamma=5\cdot 10^{-4}$. For the Landau Zener probability we observe strong differences depending on the sign of $\Theta$. The effect is limited to a small range of angles, i.e. $|\Theta| \lesssim \pi/10$, but is quantitatively substantial; for example, $P_{LZ}(\Theta=2.2\cdot 10^{-2}\pi) - P_{LZ}(\Theta=-2.2\cdot 10^{-2}\pi) \gtrsim 0.2$. Slightly smaller differences are found in the same limited angle range for the excited state survival probability. These differences are present generally at smaller sweep speeds compared to the Landau Zener probability.

\section{Crossing Time windows for relaxation}

Finding converged numerical results for the Landau Zener dynamics (as presented in the previous section) shows two important points. Firstly, convergence regarding the protocol simulation time $t_{\rm max}$ highlights that relaxation acts only in a finite time window around the avoided crossing at $t=0$. Secondly, we determine system-bath coupling strength regimes where the NEQB accurately describe the dynamics. This allows us to rationalize the crossing time window for relaxation by the time (driving) dependent relaxation rate of the NEQB.

\begin{figure}[t]
\includegraphics[width=8.5cm]{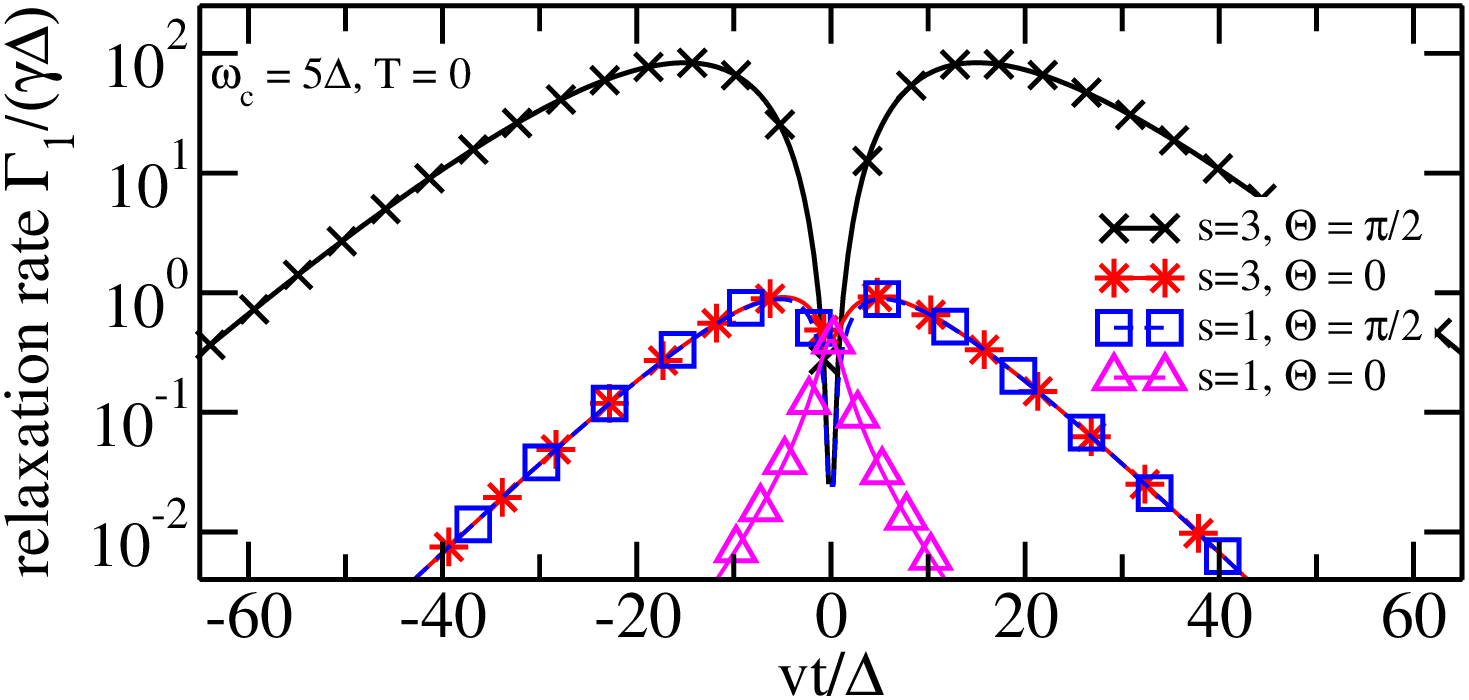}
\caption{\label{Fig5Nal} The relaxation rate $\Gamma_1[s,\Theta](t)$ (in units of $\gamma\Delta$) versus time at zero temperature for Ohmic and super-Ohmic longitudinal and transversal baths.}
\end{figure}

Fig.\ref{Fig5Nal} plots the relaxation rate $\Gamma_1[s,\Theta](t)$ for longitudinal (star and triangle symbols) and transversal (cross and square symbols)  Ohmic (square and triangle symbols) as well as super Ohmic (cross and star symbols) baths. For a longitudinal Ohmic bath (magenta triangles) the relaxation rate is maximal at the avoided crossing $t=0$ and decreases strongly on both sides of the crossing with $\propto \exp(-E(t)/\omega_c) / E(t)$. For a transversal Ohmic bath the relaxation rate vanishes at $t=0$ but quickly increases away from the crossing due to $\propto (vt)^2 \exp(-E(t)/\omega_c) /E(t) $. Finally the cut-off frequency of the environment strongly suppresses relaxation further away from the crossing. A longitudinal super-Ohmic bath exhibits an almost identical relaxation rate $\propto E(t) \exp(-E(t)/\omega_c) $ away from the crossing (where $E(t)\simeq (vt)$) but in contrast does not vanish at the crossing. A transversal super-Ohmic bath leads to a vanishing relaxation rate at $t=0$ but increases even stronger away from the crossing, i.e. $\propto (vt)^2 E(t) \exp(-E(t)/\omega_c) $. Despite the large differences of the time dependent relaxation rate in these four bath cases, convergence in all cases is finally ensured by the cut-off frequency of the bath.  

The quantitative differences for the various bath cases and, accordingly, the relevant time windows in which relaxation will influence the Landau-Zener dynamics dominantly, can be estimated from Fig.\ref{Fig5Nal}. In all cases the time window is centered around the avoided crossing but only for a longitudinal Ohmic bath it actually peaks at the crossing. The relaxation rate in the other three cases has its peaks when the energy splitting of the TSS equals for $s=1$ with $\Theta=\pi/2$ and $s=3$ with $\Theta=0$ roughly five times and for $s=3$ with $\Theta=\pi/2$ 15 times the bath cut-off energy. To compare the overall influence of relaxation in the four cases we determine 
\[ 
\Xi[s,\Theta]=\int_{-\infty}^\infty dt \Gamma_1[s,\Theta](t)
\]
and find $\Xi[1,0]=3.5 \Delta^2/(v\gamma)$ for longitudinal Ohmic relaxation, $\Xi[1,\pi/2]=47.8 \Delta^2/(v\gamma)$ for transversal Ohmic relaxation, $\Xi[3,0]=51.3 \Delta^2/(v\gamma)$ for longitudinal super-Ohmic relaxation and $\Xi[3,\pi/2]=7475 \Delta^2/(v\gamma)$ for transversal super-Ohmic relaxation. The ratios of these values reflect nicely the necessary ratios of system-bath coupling strengths to obtain similar Landau-Zener dynamics (similar values for $v_{\rm min}$) for the different bath cases.

\begin{figure}[t]
\includegraphics[width=8.5cm]{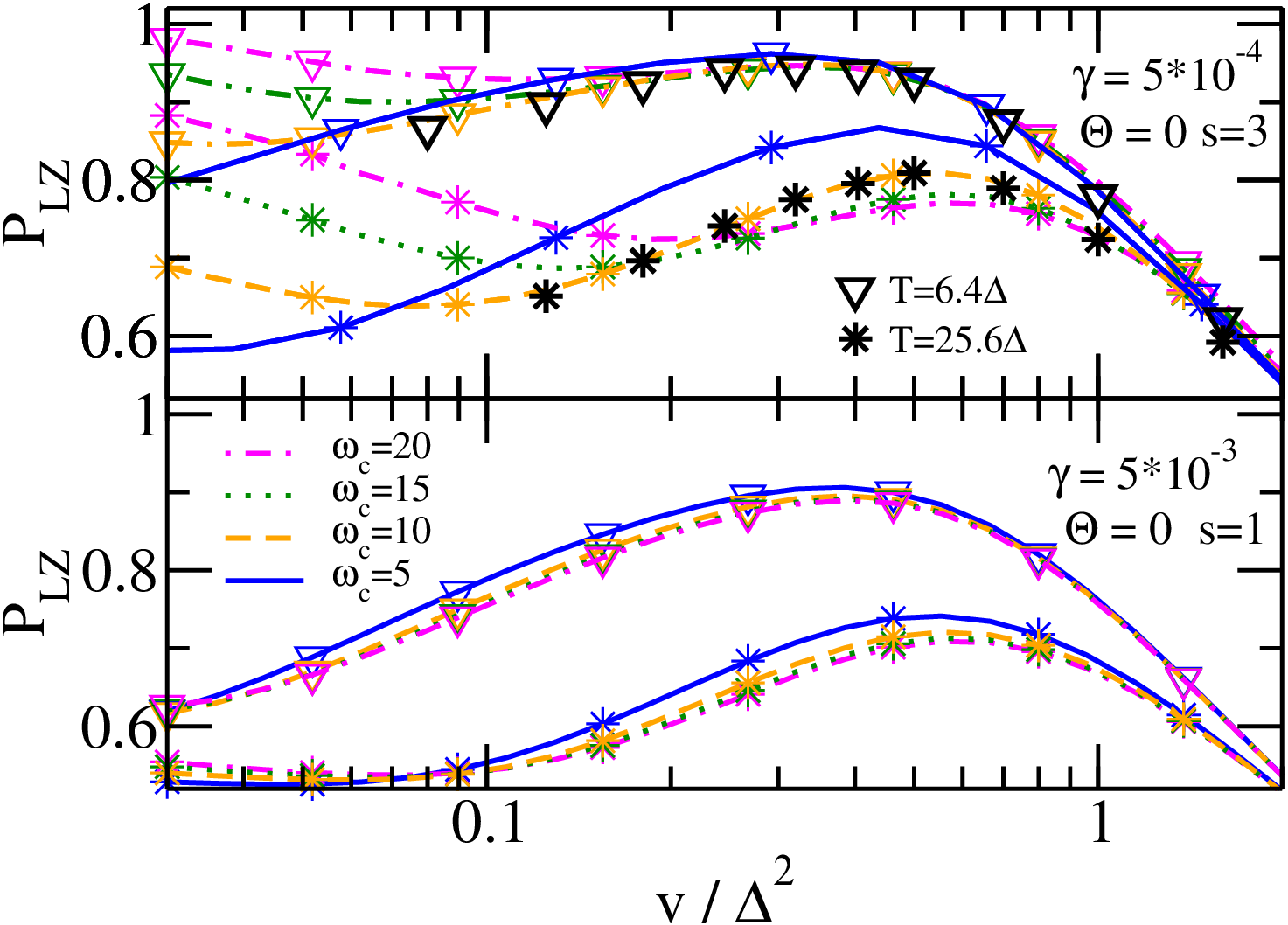}\caption{\label{Fig6Nal} The Landau-Zener probability $P_{\rm LZ}$ versus sweep speed for two temperatures $T=6.4\Delta\kb$ and $T=25.6\Delta/\kb$, various bath cut-off frequencies $\omega_c$ and (upper figure) a longitudinal super Ohmic bath with $\gamma=5\cdot 10^{-4}$ and (lower figure) a longitudinal Ohmic bath with $\gamma=5\cdot 10^{-3}$.}
\end{figure}

The vastly different crossing time windows depend in all cases strongly on $\omega_c$ except for a longitudinal Ohmic bath. This is reflected in a strong dependence of the Landau-Zener dynamics at $\omega_c$. Fig. \ref{Fig6Nal} plots the Landau Zener probability $P_{\rm LZ}$ versus sweep speed at two temperatures $T=6.4\Delta\kb$ and $T=25.6\Delta/\kb$. The upper figure represents data of a longitudinal super Ohmic bath with $\gamma=5\cdot 10^{-4}$ for four different cut-off frequencies $\omega_c$. The black symbols are QUAPI data for $\omega_c=10\Delta$. All other data is determined by NEQB. Clearly a very strong dependence on $\omega_c$ is observed within the adiabatic regime $v\lesssim \Delta^2$. In contrast a longitudinal Ohmic bath (data depicted for $\gamma=5\cdot 10^{-3}$ in the lower figure of Fig. \ref{Fig6Nal}) shows only a very weak dependence on $\omega_c$.

\section{Results}

Taking dissipation into account for a Landau-Zener driven two-state system, one might expect for any finite temperature $T$ the system to end up in its ground state since, firstly, the energy splitting of the TSS is linearly increasing and, thus, thermal equilibrium dictates occupation of the ground state. Secondly, the drive protocoll runs for infinity, i.e. $t\to\infty$, giving the system enough time to relax towards equilibrium. 

Relaxation is, however, only relevant in a finite crossing time window centered around the avoided crossing. This has been shown for Ohmic bath previously. We extended the study of dissipative Landau-Zener dynamics towards the influence of super Ohmic environments. We find finite crossing time windows. We observe that a longitudinal super Ohmic bath shows a crossing time window similar to a transversal Ohmic bath which exhibits a much wider crossing window than the longitudinal Ohmic bath. A transverse super Ohmic bath shows again a much wider crossing time window which renders a numerical exact simulation even at very small system-bath couplings challenging.

We determine the Landau-Zener and the excited state survival probabilities for a driven TSS in a super Ohmic environments numerically exact employing the quasi adiabatic path integral. We show that at weak coupling the adiabatic Markovian non-equilibrium Bloch equations can reliably simulate both. We rationalize the crossing time windows by the drive dependent weak coupling relaxation rates of the NEQBs. The crossing time windows are qualitatively determined by the bath spectrum, i.e. the cut-off frequency but large quantitative differences emerge for the studied cases in dependence of the angle $\Theta$.
Our results emphasize that relaxation dominates the dissipative influence. Dephasing, which is not expected to have a finite window of relevance, seems not to affect the studied Landau-Zener and excited state survival probabilities.

Our findings provide essential insights in the simulation of driven quantum two-state systems relevant to most future quantum electronic devices. All current proposals suffer from environmental dissipative influence and often a detailed analysis of the dominant environment is lacking. We have extended studies focussing on Ohmic environments towards super Ohmic ones. The observed finite crossing time windows show that finite driving protocols for a quantum two-state system can adequately be simulated and understood once the driving protocol pushed the two-state system out of the spectral range of the environment. Our results allow to quantitatively estimate, how far, the TSS must be driven to avoid further relaxation to influence the device.


\begin{thebibliography}{40}
\expandafter\ifx\csname natexlab\endcsname\relax\def\natexlab#1{#1}\fi
\expandafter\ifx\csname bibnamefont\endcsname\relax
  \def\bibnamefont#1{#1}\fi
\expandafter\ifx\csname bibfnamefont\endcsname\relax
  \def\bibfnamefont#1{#1}\fi
\expandafter\ifx\csname citenamefont\endcsname\relax
  \def\citenamefont#1{#1}\fi
\expandafter\ifx\csname url\endcsname\relax
  \def\url#1{\texttt{#1}}\fi
\expandafter\ifx\csname urlprefix\endcsname\relax\def\urlprefix{URL }\fi
\providecommand{\bibinfo}[2]{#2}
\providecommand{\eprint}[2][]{\url{#2}}

\bibitem[{\citenamefont{Weiss}(1998)}]{WeissBuch}
\bibinfo{author}{\bibfnamefont{U.}~\bibnamefont{Weiss}},
  \emph{\bibinfo{title}{Quantum Dissipative Systems}}
  (\bibinfo{publisher}{World Scientific}, \bibinfo{address}{Singapore},
  \bibinfo{year}{1998}), \bibinfo{edition}{2nd} ed.

\bibitem[{\citenamefont{Castelvecchi}(2017)}]{Castelvecchi2017}
\bibinfo{author}{\bibfnamefont{D.}~\bibnamefont{Castelvecchi}},
  \bibinfo{journal}{Nature} \textbf{\bibinfo{volume}{541}}, \bibinfo{pages}{9}
  (\bibinfo{year}{2017}).

\bibitem[{\citenamefont{Herbschleb et~al.}(2019)\citenamefont{Herbschleb, Kato,
  Maruyama, Danjo, Makino, Yamasaki, Ohki, Hayashi, Morishita, Fujiwara
  et~al.}}]{Herbschleb2019}
\bibinfo{author}{\bibfnamefont{E.~D.} \bibnamefont{Herbschleb}},
  \bibinfo{author}{\bibfnamefont{H.}~\bibnamefont{Kato}},
  \bibinfo{author}{\bibfnamefont{Y.}~\bibnamefont{Maruyama}},
  \bibinfo{author}{\bibfnamefont{T.}~\bibnamefont{Danjo}},
  \bibinfo{author}{\bibfnamefont{T.}~\bibnamefont{Makino}},
  \bibinfo{author}{\bibfnamefont{S.}~\bibnamefont{Yamasaki}},
  \bibinfo{author}{\bibfnamefont{I.}~\bibnamefont{Ohki}},
  \bibinfo{author}{\bibfnamefont{K.}~\bibnamefont{Hayashi}},
  \bibinfo{author}{\bibfnamefont{H.}~\bibnamefont{Morishita}},
  \bibinfo{author}{\bibfnamefont{M.}~\bibnamefont{Fujiwara}},
  \bibnamefont{et~al.}, \bibinfo{journal}{Nature Communications}
  \textbf{\bibinfo{volume}{10}} (\bibinfo{year}{2019}).

\bibitem[{\citenamefont{Miao et~al.}(2020)\citenamefont{Miao, Blanton,
  Anderson, Bourassa, Crook, Wolfowicz, Abe, Ohshima, and
  Awschalom}}]{Miao2020}
\bibinfo{author}{\bibfnamefont{K.~C.} \bibnamefont{Miao}},
  \bibinfo{author}{\bibfnamefont{J.~P.} \bibnamefont{Blanton}},
  \bibinfo{author}{\bibfnamefont{C.~P.} \bibnamefont{Anderson}},
  \bibinfo{author}{\bibfnamefont{A.}~\bibnamefont{Bourassa}},
  \bibinfo{author}{\bibfnamefont{A.~L.} \bibnamefont{Crook}},
  \bibinfo{author}{\bibfnamefont{G.}~\bibnamefont{Wolfowicz}},
  \bibinfo{author}{\bibfnamefont{H.}~\bibnamefont{Abe}},
  \bibinfo{author}{\bibfnamefont{T.}~\bibnamefont{Ohshima}}, \bibnamefont{and}
  \bibinfo{author}{\bibfnamefont{D.~D.} \bibnamefont{Awschalom}},
  \bibinfo{journal}{Science} \textbf{\bibinfo{volume}{369}},
  \bibinfo{pages}{1493} (\bibinfo{year}{2020}).

\bibitem[{\citenamefont{Keck et~al.}(2017)\citenamefont{Keck, Montangero,
  Santoro, Fazio, and Rossini}}]{KeckNJP2017}
\bibinfo{author}{\bibfnamefont{M.}~\bibnamefont{Keck}},
  \bibinfo{author}{\bibfnamefont{S.}~\bibnamefont{Montangero}},
  \bibinfo{author}{\bibfnamefont{G.~E.} \bibnamefont{Santoro}},
  \bibinfo{author}{\bibfnamefont{R.}~\bibnamefont{Fazio}}, \bibnamefont{and}
  \bibinfo{author}{\bibfnamefont{D.}~\bibnamefont{Rossini}},
  \bibinfo{journal}{New Journal of Physics} \textbf{\bibinfo{volume}{19}},
  \bibinfo{pages}{113029} (\bibinfo{year}{2017}).

\bibitem[{\citenamefont{Carrega et~al.}(2020)\citenamefont{Carrega, Crescente,
  Ferraro, and Sassetti}}]{CarregaNJP2020}
\bibinfo{author}{\bibfnamefont{M.}~\bibnamefont{Carrega}},
  \bibinfo{author}{\bibfnamefont{A.}~\bibnamefont{Crescente}},
  \bibinfo{author}{\bibfnamefont{D.}~\bibnamefont{Ferraro}}, \bibnamefont{and}
  \bibinfo{author}{\bibfnamefont{M.}~\bibnamefont{Sassetti}},
  \bibinfo{journal}{New Journal of Physics} \textbf{\bibinfo{volume}{22}},
  \bibinfo{pages}{083085} (\bibinfo{year}{2020}).

\bibitem[{\citenamefont{Ginzel et~al.}(2020)\citenamefont{Ginzel, Mills, Petta,
  and Burkard}}]{BurkardPRB2020}
\bibinfo{author}{\bibfnamefont{F.}~\bibnamefont{Ginzel}},
  \bibinfo{author}{\bibfnamefont{A.~R.} \bibnamefont{Mills}},
  \bibinfo{author}{\bibfnamefont{J.~R.} \bibnamefont{Petta}}, \bibnamefont{and}
  \bibinfo{author}{\bibfnamefont{G.}~\bibnamefont{Burkard}},
  \bibinfo{journal}{Phys. Rev. B} \textbf{\bibinfo{volume}{102}},
  \bibinfo{pages}{195418} (\bibinfo{year}{2020}).

\bibitem[{\citenamefont{Landau}(1932)}]{Landau1932}
\bibinfo{author}{\bibfnamefont{L.~D.~a.} \bibnamefont{Landau}},
  \bibinfo{journal}{Phys. Z. Sowjetunion} \textbf{\bibinfo{volume}{2}},
  \bibinfo{pages}{46} (\bibinfo{year}{1932}).

\bibitem[{\citenamefont{Zener}(1932)}]{Zener1932}
\bibinfo{author}{\bibfnamefont{C.}~\bibnamefont{Zener}},
  \bibinfo{journal}{Proc. Roy. Soc. London A} \textbf{\bibinfo{volume}{137}},
  \bibinfo{pages}{696} (\bibinfo{year}{1932}).

\bibitem[{\citenamefont{Majorana}(1932)}]{Majorana1932}
\bibinfo{author}{\bibfnamefont{E.}~\bibnamefont{Majorana}},
  \bibinfo{journal}{Nuovo Cimento} \textbf{\bibinfo{volume}{9}},
  \bibinfo{pages}{43} (\bibinfo{year}{1932}).

\bibitem[{\citenamefont{Stueckelberg}(1932)}]{Stueckelberg1932}
\bibinfo{author}{\bibfnamefont{E.~C.~G.} \bibnamefont{Stueckelberg}},
  \bibinfo{journal}{Helv. Phys. Acta} \textbf{\bibinfo{volume}{5}},
  \bibinfo{pages}{369} (\bibinfo{year}{1932}).

\bibitem[{\citenamefont{Kayanuma}(1984{\natexlab{a}})}]{KayanumaLZ1984a}
\bibinfo{author}{\bibfnamefont{Y.}~\bibnamefont{Kayanuma}},
  \bibinfo{journal}{J. Phys. Soc. Japan} \textbf{\bibinfo{volume}{53}},
  \bibinfo{pages}{108} (\bibinfo{year}{1984}{\natexlab{a}}).

\bibitem[{\citenamefont{Kayanuma}(1984{\natexlab{b}})}]{KayanumaLZ1984b}
\bibinfo{author}{\bibfnamefont{Y.}~\bibnamefont{Kayanuma}},
  \bibinfo{journal}{J. Phys. Soc. Japan} \textbf{\bibinfo{volume}{53}},
  \bibinfo{pages}{118} (\bibinfo{year}{1984}{\natexlab{b}}).

\bibitem[{\citenamefont{Ao and Rammer}(1989)}]{AoLZ1989}
\bibinfo{author}{\bibfnamefont{P.}~\bibnamefont{Ao}} \bibnamefont{and}
  \bibinfo{author}{\bibfnamefont{J.}~\bibnamefont{Rammer}},
  \bibinfo{journal}{Phys. Rev. Lett.} \textbf{\bibinfo{volume}{62}},
  \bibinfo{pages}{3004} (\bibinfo{year}{1989}).

\bibitem[{\citenamefont{Kayanuma and Nakayama}(1998)}]{KayanumaLZ1998}
\bibinfo{author}{\bibfnamefont{Y.}~\bibnamefont{Kayanuma}} \bibnamefont{and}
  \bibinfo{author}{\bibfnamefont{H.}~\bibnamefont{Nakayama}},
  \bibinfo{journal}{Phys. Rev. B} \textbf{\bibinfo{volume}{57}},
  \bibinfo{pages}{13099} (\bibinfo{year}{1998}).

\bibitem[{\citenamefont{Wubs et~al.}(2006)\citenamefont{Wubs, Saito, Kohler,
  H\"anggi, and Kayanuma}}]{WubsLZ2006}
\bibinfo{author}{\bibfnamefont{M.}~\bibnamefont{Wubs}},
  \bibinfo{author}{\bibfnamefont{K.}~\bibnamefont{Saito}},
  \bibinfo{author}{\bibfnamefont{S.}~\bibnamefont{Kohler}},
  \bibinfo{author}{\bibfnamefont{P.}~\bibnamefont{H\"anggi}}, \bibnamefont{and}
  \bibinfo{author}{\bibfnamefont{Y.}~\bibnamefont{Kayanuma}},
  \bibinfo{journal}{Phys. Rev. Lett.} \textbf{\bibinfo{volume}{97}},
  \bibinfo{pages}{200404} (\bibinfo{year}{2006}).

\bibitem[{\citenamefont{Pokrovsky and Sun}(2007)}]{PokrovskyLZ2007}
\bibinfo{author}{\bibfnamefont{V.~L.} \bibnamefont{Pokrovsky}}
  \bibnamefont{and} \bibinfo{author}{\bibfnamefont{D.}~\bibnamefont{Sun}},
  \bibinfo{journal}{Phys. Rev. B} \textbf{\bibinfo{volume}{76}},
  \bibinfo{pages}{024310} (\bibinfo{year}{2007}).

\bibitem[{\citenamefont{Nalbach and Thorwart}(2009)}]{NalbachLZ2009}
\bibinfo{author}{\bibfnamefont{P.}~\bibnamefont{Nalbach}} \bibnamefont{and}
  \bibinfo{author}{\bibfnamefont{M.}~\bibnamefont{Thorwart}},
  \bibinfo{journal}{Phys. Rev. Lett.} \textbf{\bibinfo{volume}{103}},
  \bibinfo{pages}{220401} (\bibinfo{year}{2009}).

\bibitem[{\citenamefont{Nalbach and Thorwart}(2010)}]{NalbachLZ2010}
\bibinfo{author}{\bibfnamefont{P.}~\bibnamefont{Nalbach}} \bibnamefont{and}
  \bibinfo{author}{\bibfnamefont{M.}~\bibnamefont{Thorwart}},
  \bibinfo{journal}{Chem. Phys.} \textbf{\bibinfo{volume}{375}},
  \bibinfo{pages}{234 } (\bibinfo{year}{2010}).

\bibitem[{\citenamefont{Javanbakht et~al.}(2015)\citenamefont{Javanbakht,
  Nalbach, and Thorwart}}]{NalbachLZ2015}
\bibinfo{author}{\bibfnamefont{S.}~\bibnamefont{Javanbakht}},
  \bibinfo{author}{\bibfnamefont{P.}~\bibnamefont{Nalbach}}, \bibnamefont{and}
  \bibinfo{author}{\bibfnamefont{M.}~\bibnamefont{Thorwart}},
  \bibinfo{journal}{Phys. Rev. A} \textbf{\bibinfo{volume}{91}},
  \bibinfo{pages}{052103} (\bibinfo{year}{2015}).

\bibitem[{\citenamefont{Whitney et~al.}(2011)\citenamefont{Whitney, Clusel, and
  Ziman}}]{WhitneyPRL2011}
\bibinfo{author}{\bibfnamefont{R.~S.} \bibnamefont{Whitney}},
  \bibinfo{author}{\bibfnamefont{M.}~\bibnamefont{Clusel}}, \bibnamefont{and}
  \bibinfo{author}{\bibfnamefont{T.}~\bibnamefont{Ziman}},
  \bibinfo{journal}{Phys. Rev. Lett.} \textbf{\bibinfo{volume}{107}},
  \bibinfo{pages}{210402} (\bibinfo{year}{2011}).

\bibitem[{\citenamefont{Huang and Zhao}(2018)}]{ZhaoPRA2018}
\bibinfo{author}{\bibfnamefont{Z.}~\bibnamefont{Huang}} \bibnamefont{and}
  \bibinfo{author}{\bibfnamefont{Y.}~\bibnamefont{Zhao}},
  \bibinfo{journal}{Phys. Rev. A} \textbf{\bibinfo{volume}{97}},
  \bibinfo{pages}{013803} (\bibinfo{year}{2018}).

\bibitem[{\citenamefont{Malla and Raikh}(2018)}]{RaikhPRB2018}
\bibinfo{author}{\bibfnamefont{R.~K.} \bibnamefont{Malla}} \bibnamefont{and}
  \bibinfo{author}{\bibfnamefont{M.~E.} \bibnamefont{Raikh}},
  \bibinfo{journal}{Phys. Rev. B} \textbf{\bibinfo{volume}{97}},
  \bibinfo{pages}{035428} (\bibinfo{year}{2018}).

\bibitem[{\citenamefont{Chen}(2020)}]{ChenPRB2020}
\bibinfo{author}{\bibfnamefont{R.}~\bibnamefont{Chen}}, \bibinfo{journal}{Phys.
  Rev. B} \textbf{\bibinfo{volume}{101}}, \bibinfo{pages}{125426}
  (\bibinfo{year}{2020}).

\bibitem[{\citenamefont{Makri and
  Makarov}(1995{\natexlab{a}})}]{MakriQUAPI1995a}
\bibinfo{author}{\bibfnamefont{N.}~\bibnamefont{Makri}} \bibnamefont{and}
  \bibinfo{author}{\bibfnamefont{D.~E.} \bibnamefont{Makarov}},
  \bibinfo{journal}{J. Chem. Phys.} \textbf{\bibinfo{volume}{102}},
  \bibinfo{pages}{4600} (\bibinfo{year}{1995}{\natexlab{a}}).

\bibitem[{\citenamefont{Makri and
  Makarov}(1995{\natexlab{b}})}]{MakriQUAPI1995b}
\bibinfo{author}{\bibfnamefont{N.}~\bibnamefont{Makri}} \bibnamefont{and}
  \bibinfo{author}{\bibfnamefont{D.~E.} \bibnamefont{Makarov}},
  \bibinfo{journal}{J. Chem. Phys.} \textbf{\bibinfo{volume}{102}},
  \bibinfo{pages}{4611} (\bibinfo{year}{1995}{\natexlab{b}}).

\bibitem[{\citenamefont{Palm and Nalbach}(2018)}]{PalmJCP2018}
\bibinfo{author}{\bibfnamefont{T.}~\bibnamefont{Palm}} \bibnamefont{and}
  \bibinfo{author}{\bibfnamefont{P.}~\bibnamefont{Nalbach}},
  \bibinfo{journal}{J. Chem. Phys.} \textbf{\bibinfo{volume}{149}},
  \bibinfo{pages}{214103} (\bibinfo{year}{2018}).

\bibitem[{\citenamefont{Nalbach et~al.}(2013)\citenamefont{Nalbach, Kn\"orzer,
  and Ludwig}}]{NalbachLZ2013}
\bibinfo{author}{\bibfnamefont{P.}~\bibnamefont{Nalbach}},
  \bibinfo{author}{\bibfnamefont{J.}~\bibnamefont{Kn\"orzer}},
  \bibnamefont{and} \bibinfo{author}{\bibfnamefont{S.}~\bibnamefont{Ludwig}},
  \bibinfo{journal}{Phys. Rev. B} \textbf{\bibinfo{volume}{87}},
  \bibinfo{pages}{165425} (\bibinfo{year}{2013}).

\bibitem[{\citenamefont{Nalbach}(2014)}]{NalbachLZ2014}
\bibinfo{author}{\bibfnamefont{P.}~\bibnamefont{Nalbach}},
  \bibinfo{journal}{Phys. Rev. A} \textbf{\bibinfo{volume}{90}},
  \bibinfo{pages}{042112} (\bibinfo{year}{2014}).

\bibitem[{\citenamefont{Leggett et~al.}(1987)\citenamefont{Leggett,
  Chakravarty, Dorsey, Fisher, Garg, and Zwerger}}]{Leggett1987}
\bibinfo{author}{\bibfnamefont{A.~J.} \bibnamefont{Leggett}},
  \bibinfo{author}{\bibfnamefont{S.}~\bibnamefont{Chakravarty}},
  \bibinfo{author}{\bibfnamefont{A.~T.} \bibnamefont{Dorsey}},
  \bibinfo{author}{\bibfnamefont{M.~P.~A.} \bibnamefont{Fisher}},
  \bibinfo{author}{\bibfnamefont{A.}~\bibnamefont{Garg}}, \bibnamefont{and}
  \bibinfo{author}{\bibfnamefont{W.}~\bibnamefont{Zwerger}},
  \bibinfo{journal}{Rev. Mod. Phys.} \textbf{\bibinfo{volume}{59}},
  \bibinfo{pages}{1} (\bibinfo{year}{1987}).

\bibitem[{\citenamefont{W\"urger}(1997)}]{Wuerger1997}
\bibinfo{author}{\bibfnamefont{A.}~\bibnamefont{W\"urger}},
  \bibinfo{journal}{Phys. Rev. Lett.} \textbf{\bibinfo{volume}{78}},
  \bibinfo{pages}{1759} (\bibinfo{year}{1997}).

\bibitem[{\citenamefont{W\"urger}(1998)}]{Wuerger1998}
\bibinfo{author}{\bibfnamefont{A.}~\bibnamefont{W\"urger}},
  \bibinfo{journal}{Phys. Rev. B} \textbf{\bibinfo{volume}{57}},
  \bibinfo{pages}{347} (\bibinfo{year}{1998}).

\bibitem[{\citenamefont{Nalbach}(2002)}]{NalbachRESPET2002}
\bibinfo{author}{\bibfnamefont{P.}~\bibnamefont{Nalbach}},
  \bibinfo{journal}{Phys. Rev. B} \textbf{\bibinfo{volume}{66}},
  \bibinfo{pages}{134107} (\bibinfo{year}{2002}).

\bibitem[{\citenamefont{Ludwig et~al.}(2003)\citenamefont{Ludwig, Nalbach,
  Rosenberg, and Osheroff}}]{NalbachLudwigTS2003}
\bibinfo{author}{\bibfnamefont{S.}~\bibnamefont{Ludwig}},
  \bibinfo{author}{\bibfnamefont{P.}~\bibnamefont{Nalbach}},
  \bibinfo{author}{\bibfnamefont{D.}~\bibnamefont{Rosenberg}},
  \bibnamefont{and} \bibinfo{author}{\bibfnamefont{D.}~\bibnamefont{Osheroff}},
  \bibinfo{journal}{Phys. Rev. Lett.} \textbf{\bibinfo{volume}{90}},
  \bibinfo{pages}{105501} (\bibinfo{year}{2003}).

\bibitem[{\citenamefont{Nalbach et~al.}(2004)\citenamefont{Nalbach, Osheroff,
  and Ludwig}}]{NalbachJLTPTS2004}
\bibinfo{author}{\bibfnamefont{P.}~\bibnamefont{Nalbach}},
  \bibinfo{author}{\bibfnamefont{D.~D.} \bibnamefont{Osheroff}},
  \bibnamefont{and} \bibinfo{author}{\bibfnamefont{S.}~\bibnamefont{Ludwig}},
  \bibinfo{journal}{J. of Low Temp. Phys.} \textbf{\bibinfo{volume}{137}},
  \bibinfo{pages}{395} (\bibinfo{year}{2004}).

\bibitem[{\citenamefont{Hartmann et~al.}(2000)\citenamefont{Hartmann, Goychuk,
  Grifoni, and H\"anggi}}]{HaenggiPRE2000}
\bibinfo{author}{\bibfnamefont{L.}~\bibnamefont{Hartmann}},
  \bibinfo{author}{\bibfnamefont{I.}~\bibnamefont{Goychuk}},
  \bibinfo{author}{\bibfnamefont{M.}~\bibnamefont{Grifoni}}, \bibnamefont{and}
  \bibinfo{author}{\bibfnamefont{P.}~\bibnamefont{H\"anggi}},
  \bibinfo{journal}{Phys. Rev. E} \textbf{\bibinfo{volume}{61}},
  \bibinfo{pages}{R4687} (\bibinfo{year}{2000}).

\bibitem[{\citenamefont{Solinas et~al.}(2010)\citenamefont{Solinas,
  M\"ott\"onen, Salmilehto, and Pekola}}]{PekolaPRB2010}
\bibinfo{author}{\bibfnamefont{P.}~\bibnamefont{Solinas}},
  \bibinfo{author}{\bibfnamefont{M.}~\bibnamefont{M\"ott\"onen}},
  \bibinfo{author}{\bibfnamefont{J.}~\bibnamefont{Salmilehto}},
  \bibnamefont{and} \bibinfo{author}{\bibfnamefont{J.~P.}
  \bibnamefont{Pekola}}, \bibinfo{journal}{Phys. Rev. B}
  \textbf{\bibinfo{volume}{82}}, \bibinfo{pages}{134517}
  (\bibinfo{year}{2010}).

\bibitem[{\citenamefont{Arceci et~al.}(2017)\citenamefont{Arceci, Barbarino,
  Fazio, and Santoro}}]{SantoroLZ2017}
\bibinfo{author}{\bibfnamefont{L.}~\bibnamefont{Arceci}},
  \bibinfo{author}{\bibfnamefont{S.}~\bibnamefont{Barbarino}},
  \bibinfo{author}{\bibfnamefont{R.}~\bibnamefont{Fazio}}, \bibnamefont{and}
  \bibinfo{author}{\bibfnamefont{G.~E.} \bibnamefont{Santoro}},
  \bibinfo{journal}{Phys. Rev. B} \textbf{\bibinfo{volume}{96}},
  \bibinfo{pages}{054301} (\bibinfo{year}{2017}).

\bibitem[{\citenamefont{Arceci et~al.}(2018)\citenamefont{Arceci, Barbarino,
  Fazio, and Santoro}}]{SantoroLZ2017Erratum}
\bibinfo{author}{\bibfnamefont{L.}~\bibnamefont{Arceci}},
  \bibinfo{author}{\bibfnamefont{S.}~\bibnamefont{Barbarino}},
  \bibinfo{author}{\bibfnamefont{R.}~\bibnamefont{Fazio}}, \bibnamefont{and}
  \bibinfo{author}{\bibfnamefont{G.~E.} \bibnamefont{Santoro}},
  \bibinfo{journal}{Phys. Rev. B} \textbf{\bibinfo{volume}{98}},
  \bibinfo{pages}{019902} (\bibinfo{year}{2018}).

\bibitem[{\citenamefont{Nalbach et~al.}(2017)\citenamefont{Nalbach,
  Klinkenberg, Palm, and M\"uller}}]{NalbachPRE2017}
\bibinfo{author}{\bibfnamefont{P.}~\bibnamefont{Nalbach}},
  \bibinfo{author}{\bibfnamefont{N.}~\bibnamefont{Klinkenberg}},
  \bibinfo{author}{\bibfnamefont{T.}~\bibnamefont{Palm}}, \bibnamefont{and}
  \bibinfo{author}{\bibfnamefont{N.}~\bibnamefont{M\"uller}},
  \bibinfo{journal}{Phys. Rev. E} \textbf{\bibinfo{volume}{96}},
  \bibinfo{pages}{042134} (\bibinfo{year}{2017}).

\end{thebibliography}
\end{document}